\documentclass[sigconf,authorversion]{acmart} 

\usepackage{booktabs} 
\usepackage{enumitem}

\copyrightyear{2025}
\acmYear{2025}
\setcopyright{acmlicensed}
\acmConference[SAC '25]{The 40th ACM/SIGAPP Symposium on Applied Computing}{March 31-April 4, 2025}{Catania, Italy}
\acmBooktitle{The 40th ACM/SIGAPP Symposium on Applied Computing (SAC '25), March 31-April 4, 2025, Catania, Italy}
\acmDOI{10.1145/3672608.3707855}
\acmISBN{979-8-4007-0629-5/25/03}

\usepackage{amsmath,amsfonts,amstext,latexsym}
\usepackage{algorithm,algorithmicx}
\usepackage[noend]{algpseudocode}
\usepackage{listings}
\usepackage{array}
\usepackage{microtype}
\usepackage{xparse}
\usepackage{subcaption}

\algdef{SE}[SUBALG]{bindent}{eindent}{}{\algorithmicend}
\algtext*{bindent}
\algtext*{eindent}
\newcommand{\mathsc}[1]{{\normalfont\textsc{#1}}}

\lstdefinestyle{mystyleNumber}{
    numberstyle=\tiny\color{black!50},
    breakatwhitespace=false,
    breaklines=true,
    captionpos=b,
    keepspaces=true,
    numbers=left,
    numbersep=5pt,
    showspaces=false,
    showstringspaces=false,
    showtabs=false,
    tabsize=2,
    float=t,
    belowcaptionskip=-1em,
    frame=single,
    basicstyle=\ttfamily
}

\lstdefinestyle{mystyleNoNumber}{
    backgroundcolor=\color{backcolour},
    numberstyle=\tiny\color{codegray},
    breakatwhitespace=false,
    breaklines=true,
    captionpos=b,
    keepspaces=true,
    showspaces=false,
    showstringspaces=false,
    showtabs=false,
    tabsize=2
}

\lstdefinestyle{mystyleInLine}{
    breakatwhitespace=false,
    breaklines=true,
    keepspaces=true,
    showspaces=false,
    showstringspaces=false,
    showtabs=false,
    tabsize=2
}

\lstdefinelanguage{ASM}
{keywords={rule, if, then, endif, else, forall, choose, in, do, let, par, endpar, seqblock, endseqblock, derived, function, universe, where, or, and, not, with, seq, for, iterate, by, import, endseq},
sensitive=false,
mathescape=true
}

\newcommand{\be}{\overline{\varphi}}

\NewDocumentCommand{\kpid}{o o}{
  \mathit{Kp}\IfValueT{#1}{_{#1}}, 
  \mathit{Ki}\IfValueT{#1}{_{#1}}%
  \IfValueTF{#2}{\ \text{#2}\ }{, }%
  \mathit{Kd}\IfValueT{#1}{_{#1}}
}

\newcommand{\id}{identifier} 

\newcommand{\bs}{w}   
\newcommand{\bss}{w'} 
\newcommand{\key}{r}  
\newcommand{\sz}{|w|} 
\newcommand{\nm}{|M|} 
\newcommand{\hd}{d}   
\newcommand{\sm}{s}   

\newcommand{\re}{\mathit{recoverExprs}}
\newcommand{\qp}{\mathit{queryPUF}}

\begin{document}
\title{Software-Hardware Binding for Protection of\\Sensitive Data in Embedded Software}

\renewcommand{\shorttitle}{Software-Hardware Binding for Protection of Sensitive Data in Embedded Software}

\author{Bernhard Fischer}
\orcid{0000-0001-9737-0056}
\author{Daniel Dorfmeister}
\authornote{Corresponding author: daniel.dorfmeister@scch.at}
\orcid{0000-0002-2718-6007}
\author{Flavio Ferrarotti}
\orcid{0000-0003-2278-8233}
\affiliation{%
  \institution{Software Competence Center Hagenberg}
  \city{Hagenberg}
  \country{Austria}
}

\author{Manuel Penz}
\orcid{0009-0002-4987-2507}
\author{Michael Kargl}
\orcid{0009-0009-7753-3335}
\author{Martina Zeinzinger}
\orcid{0009-0009-5471-9129}
\author{Florian Eibensteiner}
\orcid{0000-0002-5311-4082}
\affiliation{%
  \institution{University of Applied Sciences Upper Austria}
  \city{Hagenberg}
  \country{Austria}
}

\renewcommand{\shortauthors}{B. Fischer et al.}

\begin{abstract} 
Embedded software used in industrial systems frequently relies on data that ensures the correct and efficient operation of these systems.
Thus, companies invest considerable resources in fine-tuning this data, making it their valuable intellectual property (IP).
We present a novel protection mechanism for this IP that combines hardware fingerprints with Boolean logic.
Unlike usual copy-protection approaches, unauthorised copies of the software still run on cloned devices but suboptimally.
According to our security evaluation, only a complex dynamic analysis of the protected software running on the genuine target device can reveal the secret data.
This makes the protection offered by our method more difficult to bypass.
Notably, our approach does not require additional hardware, relying only on relatively simple updates to the software.
We evaluate our protection mechanism by binding the parameters of a PID controller to a microcontroller unit (MCU) by using a physically unclonable function (PUF) based on its SRAM.
\end{abstract}

%
%

\ccsdesc[500]{Security and privacy~Software and application security}
\ccsdesc[300]{Security and privacy~Software reverse engineering}
\ccsdesc[300]{Computer systems organization~Embedded systems}

\keywords{software protection, software-hardware binding, embedded software, SRAM PUF, intellectual property}

\maketitle

\pagebreak
\section{Introduction}

Embedded software is a fundamental component in a wide range of applications, from industrial automation to consumer electronics and IoT devices.
As it is so widely used, there are ever-increasing security concerns, and it becomes even more important to protect it against unauthorised access, tampering and copying.
In this paper, we focus on a key aspect of the latter.
More precisely, our goal is to protect intellectual property (IP) in embedded software.
We aim for a protection mechanism that not only prevents the use of unauthorised copies of the software on cloned hardware but also prevents attackers from learning sensitive data from it. 

Physically unclonable functions (PUFs) provide a robust method to extract fingerprints that uniquely distinguish genuine hardware from hardware clones.
PUFs are hardware-based security primitives~\cite{gassend2002silicon, herderPufTutorial, pufTaxonomyMcGrath}.
Intrinsic PUFs leverage minor unavoidable deviations in the manufacturing process of a hardware component, which lead to slightly different electrical characteristics or behaviour. 
PUFs thus provide a good base to securely bind software to hardware, so that a given program only runs correctly on the genuine target device, and does not run or runs suboptimally on cloned devices.

\begin{figure}
    \centering
    \includegraphics[width=\columnwidth]{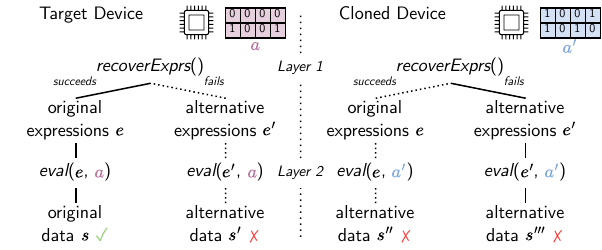}
    \caption{Dual layer protection: 
    Only if $\re$ succeeds obtaining the Boolean expressions $e$ on the target device where the PUF response provides the Boolean assignment $a$, then the secret data $s$ is recovered.
    Otherwise, i.e., whenever $\re$ fails or the PUF response gives a Boolean assignment $a' \neq a$ (or both), then some predefined alternative data (not necessarily the same in all cases) is returned.}
    \label{fig:overview}
\end{figure}

In concrete terms, we propose a new approach for software-hardware binding that protects sensitive data in embedded software, as illustrated in \autoref{fig:overview}.
The protection relies on the simple but key observation that we can encode any binary string $s$ (secret data) using a list of Boolean expressions $e$, so that an evaluation of $e$ results in $s$ only if a correct Boolean assignment $a$ is provided for the variables in $e$.
Assuming that there are $n$ different Boolean variables in the expressions in $e$, we can hide the secret data as the response to only one of $2^n$ possible assignments.
The idea is that the correct assignment $a$ is provided by a PUF response on the target device, so that the original data can be retrieved only on the target device.
On any other (cloned) device, the PUF response $a'$ will differ from $a$ and thus the evaluation of $e$ returns alternative data.
By construction of the Boolean expressions in $e$, we can precisely control the alternative data.
This latter property is very useful, as we can thus provide alternative data that degrades the performance of the system in a controlled way, which makes the protection not immediately obvious to an attacker.
In addition, it avoids the possibility of a catastrophic failure in the event of an erroneous PUF response on the target device, e.g., due to an attack on the PUF~\cite{Roelke2016}.
Think, for instance, about control systems of airplanes, cars, industrial robots or power plants.

In addition to encoding the data into Boolean expressions, we build another layer of protection by transforming $e$ into an alternative list of expressions $e'$, so that none of the Boolean assignments, not even $a$, results in $e'$ evaluating to the original data $s$.
We use a secure procedure $\re$ that returns the correct list of expressions $e$ only if certain correct credentials are provided.
In our proof of concept, the Boolean expressions are encrypted and $\re$ returns $e$ only if the correct cryptographic key is given.
Otherwise, $\re$ returns $e'$.
However, other approaches are possible here.
For instance, on a device with DRAM, we could use GlueZilla's approach~\cite{gluezilla} to directly transform $e'$ into $e$ using Rowhammer, which has similar properties to PUFs and thus only works on the target device.

We demonstrate our mechanism by protecting the proportional ($\mathit{Kp}$), integral ($\mathit{Ki}$) and derivative ($\mathit{Kd}$) parameters of a PID controller.
Regarding hardware, we use a fairly typical setup in industrial production systems (IPS) consisting of a microcontroller unit (MCU) equipped with flash memory and static random access memory (SRAM).
In this context, SRAM-based PUFs are an obvious choice.
In \autoref{sec:puf}, we present an SRAM PUF suitable to protect the PID controller so that it only runs optimally on its target MCU.
On any other hardware, it instead behaves suboptimally, even on exact replicas of the target MCU.
We apply the protection mechanism by encoding the secret data ($\kpid[][and]$ in our example) as Boolean expressions using the automated procedure described in Section~\ref{sec:mechanism}.
We use a PUF response of the target MCU to assign appropriate values to the Boolean variables in the expressions, evaluating these recovers the secret data.
Any other Boolean assignment results in incorrect (suboptimal) $\kpid[][and]$ values being returned.

Note that combining PUF responses with Boolean expressions encoding data indeed offers increased security over the classical alternative of simply encrypting the data.
All an attacker needs to break standard data encryption is to obtain the decryption key.
In our case, the data is not at all present in the protected code, not even in encrypted form.
An attacker first needs to retrieve the Boolean expressions.
Additionally, they also need to obtain the correct PUF response to decode the secret data from these Boolean expressions.
We provide a proper security evaluation of this fact in \autoref{sec:secEval}.
Moreover, our approach facilitates the application of ideas from code diversification and obfuscation~\cite{HosseinzadehRLM18} to increase security even further.
For instance, we can produce different Boolean expressions for each individual device.
Finally, dual layer protection opens up novel applications to areas such as secure update, as discussed in \autoref{sec:discussion}.

The paper is organised as follows:
In \autoref{sec:prelim}, we present the main technical concepts used throughout the paper, as well as the PID controller algorithm used in our proof of concept.
The threat model that underpins our protection approach is described in \autoref{sec:threat-model}.
The main scientific and technical contributions of our paper in Sections~\ref{sec:puf} and \ref{sec:mechanism} present the steps to enrol and query SRAM PUFs on our target hardware and our novel protection mechanism.
A detailed security evaluation, our proof of concept and its evaluation, as well as a discussion follow in Sections~\ref{sec:secEval}--\ref{sec:discussion}.
\autoref{sec:related-work} discusses related work and \autoref{sec:concl} concludes our paper.

\section{Preliminaries}\label{sec:prelim}

We consider an industrial environment where programs run on an MCU, a small computing unit integrated on a single chip with restricted resources in terms of computational power and memory size.
Consequently, we use standard STM32 Nucleo-144 development boards with STM32F767ZI MCU\footnote{For details see \url{https://www.st.com/en/evaluation-tools/nucleo-f767zi.html}}, 2 MiB of flash memory, and 512 KiB of SRAM. 
Flash and SRAM memory are the norm when it comes to MCUs' storage.
The former is where the binary programs, usually compiled C code, are stored. 
The latter is where dynamic data used by the programs is read and written during execution.

\paragraph{SRAM PUFs}
Although to date, over 40 different approaches for the realisation of PUFs~\cite{pufTaxonomyMcGrath} have been proposed, it is advantageous to implement them with readily available hardware components such as SRAM.
SRAM PUFs are based on the initial values of memory cells when powering on the memory.
Manufacturing variations affect the transistors in each cell.
These transistors have minimal differences in their threshold voltage, which determines the initial state of the cell. 
Thus, powering on any given SRAM chip causes it to be initialised with a random but unique bit pattern.
Furthermore, the values 0 and 1 are equally distributed in the pattern~\cite{herderPufTutorial}.

A PUF must reliably produce the same result with every query on the same device, and the PUF response must be clearly distinguishable between different devices.
Since SRAM initialisation is susceptible to noise, it cannot be used directly as a unique \id{} of a specific SRAM chip or MCU.
\citeauthor{Zeinzinger2023}~\cite{Zeinzinger2023} showed that the extent of this SRAM initialisation noise varies between devices and even more so between different models and that it especially depends on ambient factors such as temperature.
However, they also showed that a large percentage of SRAM cells behaves consistently on every startup, making SRAM suitable for the implementation of a PUF.
PUFs that are prone to noise are commonly processed using fuzzy extractors.
\citeauthor{Canetti2020} proposed such a fuzzy extractor~\cite{Canetti2020} based on digital lockers~\cite{Canetti2008}.

\paragraph{PID Controllers}
We illustrate our protection mechanism by considering a generic proportional-integral-derivative controller (PID controller).
They are used for a wide range of industrial applications.
For example, they can be used to control the position of a DC motor shaft.
The abstract state machine (ASM) rule in \autoref{lst:GenericPIDcontroller} specifies the step-by-step behaviour of a generic PID controller algorithm.
The ASM method is a well-known formalism for the specification of programs, see, e.g.,~\cite{BorgerR18, boerger:2003}.
Readers not familiar with ASMs can interpret ASM rules as pseudocode that produces a set of updates to the current state.
An ASM applies its rules until no more updates are produced.

We assume the initial state of $\mathit{mode} = \texttt{init}$ and that the time interval between controlling actions is $\mathit{dt}$.
Note that $\mathit{Kp}$, $\mathit{Ki}$ and $\mathit{Kd}$ take constant values.
They depend on the application at hand and are typically defined by engineers through trial and error, as there is no general analytical method to determine them. 
For complex controllers, finding optimal $\mathit{Kp}$, $\mathit{Ki}$ and $\mathit{Kd}$ parameters is not trivial. 
Indeed, this is the type of IP that we want to protect.

\begin{lstlisting}[language=ASM, style=mystyleNumber, caption={Generic PID Controller}, label = lst:GenericPIDcontroller]
$\mathsc{PIDController}=$
if $\mathit{mode}$ = init then
  $\mathit{previousIntegral} := 0$
  $\mathit{previousError} := 0$
  $\mathit{Kp} := 800$
  $\mathit{Ki} := 1000$
  $\mathit{Kd} := 30$
  $\mathit{mode}$ := execute
if $\mathit{mode}$ = execute then
  if $\mathit{passed}(\mathit{dt})$ then
    let $\mathit{error} = \mathit{setPoint} - \mathit{measuredValue}$ in
    let $\mathit{p} = \mathit{error}$ in
    let $\mathit{i} = \mathit{previousIntegral} + \mathit{error} \times \mathit{dt}$ in
    let $\mathit{d} = (\mathit{error} - \mathit{previousError}) \, / \, \mathit{dt}$ in
    let $\mathit{output} = \mathit{Kp} \cdot \mathit{p} + \mathit{Ki} \cdot \mathit{i} + \mathit{Kd} \cdot \mathit{d}$ in 
      if $\mathit{safeLowerRange} \leq \mathit{output} \leq \mathit{safeUpperRange}$ then 
        $\mathit{adjustControlTo}(\mathit{output})$
      if $\mathit{safeLowerRange} > \mathit{output}$ then 
        $\mathit{adjustControlTo}(\mathit{safeLowerRange})$
      if $\mathit{output} > \mathit{safeUpperRange}$ then 
        $\mathit{adjustControlTo}(\mathit{safeUpperRange})$
      $\mathit{previousIntegral} := \mathit{i}$
      $\mathit{previousError} := \mathit{error}$
\end{lstlisting}

\section{Threat Model}\label{sec:threat-model}

We make the following assumptions about the target hardware, the attacker, and the protected applications.

The hardware we consider are MCUs, i.e., small computers on integrated circuits, which are often used for automation in the manufacturing industry.
They have essential security features, such as read and write protection and a memory protection unit.
We want to protect critical/sensitive data used in embedded software on MCUs so that an attacker cannot clone the hardware and then simply copy its software to achieve the exact same functionality as on the genuine MCU.
Security features such as readout protection try to prevent this, but one should not solely rely on them and instead think about additional security layers.
In this work, we focus mainly on the protection of embedded software, i.e., the flash image.
How sensitive data evaluated at runtime by the software can be protected within SRAM, especially with regard to legacy systems, is future work.

Consequently, we assume a static white-box man-at-the-end attack scenario.
In this model, the attacker can get a copy of the binary image of the protected software by any means.
For example, by stealing it from the storage of a genuine target machine or intercepting software updates.
However, the attacker is unable to perform a dynamic analysis of the software on the genuine target machine associated with the software instance that they have obtained.
This is a reasonable assumption, as most MCUs provide ways of disabling or blocking the debugging and communication interfaces.
In addition, it is only possible to intercept the buses of the main components such as the processor and SRAM with considerable physical intervention, such as chip scraping, since these are contained on a single chip.
It has been shown that MCUs' firmware protection can be defeated by a sufficiently motivated attacker~\cite{ObermaierT17}, but this is outside the scope of our work.
The attacker in our model can create a cloned instance of the target MCU.
Therefore, a dynamic analysis of the protected software is still possible on that cloned MCU.

Attackers' goal is to create copies of the protected software that exactly perform the intended actions when deployed on any clone of the target MCU.
They do not have access to the original source code but can use a decompiler to obtain a high-level program corresponding to the protected version.

\section{SRAM PUF}
\label{sec:puf}

We use Yagemann's implementation\footnote{See \url{https://github.com/carter-yagemann/python-fuzzy-extractor}} of the concepts of \citeauthor{Canetti2020}~\cite{Canetti2020} as basis for our SRAM PUF implementation, see also~\cite{Penz2025}.
Like \citeauthor{Canetti2020}, we sample specific subsets of the bit string $\bs$ representing the PUF area, which contains $\sz$ bits.
We use these sampled subsets to encrypt an arbitrary predetermined \id{} $\key$ that represents the device, resulting in a ciphertext.
To reproduce $\key$, we use the same subset of the bit string $\bss$ to decrypt the ciphertext.
If $\bss$ represents the same PUF area during decryption, this will lead to the subsets containing the same bits and therefore the \id{} $\key$ being reconstructed.
If we take $\bss$ from another PUF area or the same area on a different device, the sampled subsets will show a different bit pattern and $\key$ cannot be successfully decrypted.
Due to the fact that only one of the subsets of $\bs$ and $\bss$ must match, the reconstruction of $\key$ is resistant to sporadic noise in $\bss$.
We thus use $\key$ as error-corrected PUF response in our protection mechanism.

The major difference between our implementation and that of \citeauthor{Canetti2020} is that in their method, the subsets of $\bs$ are sampled randomly, whereas in our approach, the subsets are sampled deterministically.
This sampling of subsets of $\bs$ can be done by masking out all indices that should not be included in the sampled subset.

We aim for a high probability of correctly decrypting the \id{} $\key$, while not being susceptible to attacks that leverage the masks $M$ with the highest number of bits that are zero.
To achieve this, \citeauthor{Canetti2020} must ensure that both the required bit string size $\sz$ and the number of masks $\nm$ are relatively high.
Thus, they exceed the memory constraints of typical MCUs.
In our concept, we construct the masks based on the maximum Hamming distance $\hd$ between the bit strings $\bs$ and $\bss$ so that $\bss$ is still able to decrypt $\key$.
We create all possible masks with a Hamming distance $\hd$ from a bit string with length $\sz$ containing only ones instead of randomly generated masks.
This approach ensures that no SRAM bit string $\bss$ with a Hamming distance greater than $\hd$ from $\bs$ can decrypt $\key$.
Thus, it is impossible for a mask to have more than $\hd$ zeros, mitigating the risk that comes with masks containing many zeros.
This is important as the number of zeros in a mask is inversely related to the time required to guess a viable substitute $\bss$ that can decrypt $\key$ without having any knowledge about the actual PUF area.

The major drawback of this method is the problem of an exploding number of masks when increasing the allowed Hamming distance $\hd$ between $\bs$ and $\bss$.
The number of masks $\nm$ depends on the Hamming distance $\hd$ and the bit string size $\sz$:
\begin{align*} 
  \nm = \frac{\sz!}{(\sz-\hd)!}
\end{align*}

With an increase in $\hd$, $\nm$ increases rapidly.
Thus, the SRAM area containing the bit string $\bs$ has to be specifically selected to produce results that are as similar as possible between multiple startups.
Long-term stability in the bit string $\bs$ enables us to reduce the allowed Hamming distance $\hd$, since our concept is then only needed to compensate for sporadic errors in the bit string $\bss$.

We define a \textit{stable bit} as a bit that has the same value over at least 99.9\% of all recorded values for this bit after the startup of the SRAM.
To determine the stable bits, we have to restart the SRAM multiple times, record its content on every startup and compare the contents at different startups.
Therefore, we need numerous initial samples to assess whether an area in the SRAM is reliable enough for our approach.
Since the stability of the initial value of the SRAM cells also depends on the temperature~\cite{Zeinzinger2023}, we preferably draw these samples in different temperature ranges.
To gather a sufficiently large number of samples with reasonably limited human interaction required, we have devised a largely automated enrolment process.

The enrolment process consists of several steps:
The sampling at the very beginning happens multiple times and ideally for different temperature ranges.
For each sample, the board under test needs to be powered down before a new measurement can take place.
Because the initial state of the SRAM is subject to a certain amount of noise, it is not always possible to find an ideal address range of the required size $\sz$ where all bits are stable across the readings taken from it.
As a remedy, we mask out individual unstable bits within this range using a device-specific \textit{stability mask} $\sm$.
The resulting bit string $\bs$, with all unstable bits masked out, is the reference value of our PUF.
Henceforth, we use the term \textit{helper data} to refer to the address range for the PUF bit string, the stability mask $\sm$, and the parameters $\sz$, $\nm$ and $\hd$ defined above for creating the set of subset sampling masks and the set of ciphertexts.
In the next step, we verify that the \id{} $\key$ can be reconstructed under various conditions using this helper data.
If the verification fails more often than specified by a cut-off, we deem the PUF unreliable.
There are several options to reduce the decryption error rate, e.g., increasing the number of samples, extending the temperature range in which we take samples, searching for a more stable address range, and adjusting the parameters $\sz$ or $\hd$.
If the verification step still fails despite these adjustments, we consider the chip unsuitable for our approach.

In order to decrypt $\key$ on a running MCU, we perform the following steps:
We read the defined address range to get the PUF's basic bit string from the SRAM, which is only possible during the startup phase.
Next, we mask out the unstable bits by applying the stability mask $\sm$, resulting in $\bss$.
We decrypt the \id{} $\key$ using the fuzzy extractor with the helper data generated during enrolment.
This is only possible if $\bss$ is within the Hamming distance $\hd$ of $\bs$.
We obfuscate the underlying PUF area afterwards, thus the PUF's memory region is usable during runtime.

\section{Protection Mechanism}
\label{sec:mechanism}

We introduce our protection mechanism by applying it to a concrete example, i.e., to safeguard the optimal proportional, integral and derivative parameters of a PID controller.
We would like to stress that the exact same mechanism can be applied in general to protect any valuable data that appears in any given program, as long as a suitable PUF is available for the target hardware.
Indeed, it should be clear from the context that our protection mechanism is widely applicable and is not restricted in any way to similar use cases. 

Assume we would like to protect the values $\kpid[0][and]$ of a PID controller as the one shown in \autoref{lst:GenericPIDcontroller}, and that we have a list of $m > 2$ alternative values $\kpid[1], \ldots, \kpid[m]$ for the constants of the PID controller.
Here, $m$ should be chosen as large as possible.
These alternative values are suboptimal but stable (cf. the security analysis in \autoref{sec:secEval}).
That is, they converge to the desired point, but they do so in more steps than when using the optimal values.
The idea is that only these alternative values can be obtained from a copy of the protected PID controller.
Unless an attacker can get hold of the correct PUF responses from the target hardware, the actual optimal values $\kpid[0][and]$ cannot be inferred from the protected program.
Moreover, even if an attacker knows the correct PUF responses, they would still need to perform a complex dynamic analysis of the protected PID controller to reveal these values. 

Let $T = \{(\kpid[i]) \mid 0 \leq i \leq m\}$ and let $V = \{x \mid y \in T \, \text{and} \, x \, \text{appears in} \, y\}$.
Then $|V|$ is the number of different values that appear in the union of the domains of the relation $T$.
We fix an arbitrary bijective function $\mathit{toBin}: V \cup V' \rightarrow \{0,1\}^n$, where $V'$ is an arbitrary set of values of size $2^n - |V|$ such that $V' \cap V = \emptyset$. Of course, $n$ must be greater than or equal to $\lfloor \log_2 |V| \rfloor + 1$.
We abuse the notation and write $\mathit{toBin}(\kpid[i])$, where $(\kpid[i]) \in T$, to denote the binary string $\mathit{toBin}(Kp_i) \cdot \mathit{toBin}(Ki_i) \cdot \mathit{toBin}(Kd_i)$ of length $3n$.  

We further assume a procedure $\mathit{encodeExprs}$ that protects a given tuple $\be$ of Boolean expressions, so that they can be decoded securely by a function $\re$ executed with the appropriate credentials on the target hardware.
In our proof of concept, we implement $\mathit{encodeExprs}$ using encryption, where the target hardware fingerprint provides the decryption key.
Alternatives are, of course, possible, but this is not the main focus of this paper.
We let $\mathit{hashValue}$ be the necessary token to check whether $\re$ is successfully executed, i.e., $\mathit{hash}(\re(\mathit{encodeExprs}(\be)) = \mathit{hashValue}$.
Finally, we assume a function $\qp(k)$ that returns a binary substring of length $k$ of the \id{} $\key$ described in \autoref{sec:puf}.

Our copy-protection mechanism is implemented as follows:
\begin{enumerate}[leftmargin=*]
    \item Define a partition of $\{0,1\}^k$, for some $k \geq \lfloor \log_2(m+1) \rfloor + 1$, into $m+1$ nonempty subsets $A_i$, where $0 \leq i \leq m$. Recall that $m+1$ is the number of different triples of proportional, integral, and derivative values in $T$.
    \item Choose a $c$ with $1 < c \leq m$ and define $f: \{0,1\}^k \rightarrow \{0,1\}^{3n}$ and $f': \{0,1\}^k \rightarrow \{0,1\}^{3n}$ as follows: 
    \begin{align*}
    f(x) = &\mathit{toBin}(\kpid[i]) \, \text{if} \, x \in A_i.\\
    f'(x) = &\begin{cases}
        f(x), & \text{if} \, x \in A_i \, \text{for} \, 1 \leq i \leq c\\
        \mathit{toBin}(\kpid[c]), & \text{if} \, x \in A_i \, \text{for} \, i = 0 \, \text{or} \, c < i \leq m   
    \end{cases}     
    \end{align*}
    \item Build tuples of Boolean expressions $\be = (\varphi_0, \ldots, \varphi_{3n-1})$ and $\be' = (\varphi_0', \ldots, \varphi_{3n-1}')$ on Boolean variables $x_0, \ldots, x_{k-1}$ (using the procedure explained below), such that for every assignment $\mathit{val}: \{x_0, \ldots, x_{k-1}\} \rightarrow \{0, 1\}$ and  binary string $b_0 \ldots b_{3n-1} \in \{0,1\}^{3n}$, the following holds:
    \begin{align*}
        f(\mathit{val}(x_0),& \ldots, \mathit{val}(x_{k-1})) =  b_0 \ldots b_{3n-1} \\
        &\text{iff} \; \mathit{val}(\varphi_i) = b_i \, \text{for all} \, i = 0, \ldots, 3n-1
    \end{align*} 
    and 
    \begin{align*}
        f'(\mathit{val}(x_0),& \ldots, \mathit{val}(x_{k-1})) =  b_0, \ldots, b_{3n-1} \\
        &\text{iff} \; \mathit{val}(\varphi_i') = b_i \, \text{for all} \, i = 0, \ldots, 3n-1.
    \end{align*} 
    \item Set $\mathit{encodedExprs} := \mathit{encodeExprs}(\be)$.
    \item Choose $\qp(k)$ so that its response belongs to $A_0$ on the target machine.
    \item Modify the code of the PID controller by replacing the assignment of constant values to $\kpid[][and]$ (e.g., lines~5--7 in \autoref{lst:GenericPIDcontroller}) with the code in \autoref{lst:ProtectionKpKiKdValues}.
    Thus, if $\be$ recovered successfully (lines~2--4), then the algorithm evaluates each of these Boolean expressions using the $k$ bits in the response of $\qp$ as assignment for the Boolean variables $x_0, \ldots, x_{k-1}$ (line~4), where the $i$-th bit is assigned to $x_i$  for $i = 0, \ldots, k-1$.
    The function $\mathit{eval}$ returns a binary string $b_0, \ldots, b_{3n-1}$ such that $b_i$ is $1$ iff $\varphi_1$ is true for the given assignment based on the response of $\qp$.
    Then, the inverse of the function $toBin$ returns (provided  $\qp \in A_0$) the optimal values for $\kpid$.
    Otherwise, the algorithm assumes we are not on the target machine and uses the Boolean expressions $\be'$ instead.
    In this case (line~6) the values assigned to $\kpid$ will be, by construction of these Boolean expressions, among the alternative suboptimal ones.    
\end{enumerate}

\begin{lstlisting}[language=ASM, style=mystyleNumber, caption={Secure Recovery of Protected $\mathit{Kp}$, $\mathit{Ki}$ and $\mathit{Kd}$ values}, label = lst:ProtectionKpKiKdValues]
$\mathsc{RecoverProtectedValuesPID}=$
let $\be = \re(\mathit{encodedExprs})$
if $\mathit{hash}(\be) = \mathit{hashValue}$ then 
  $\smash{\kpid := \mathit{toBin}^{-1}(\mathit{eval}(\be, \qp(k)))}$
else
  $\smash{\kpid := \mathit{toBin}^{-1}(\mathit{eval}(\be', \qp(k)))}$
\end{lstlisting}

The Boolean expressions as required for Step~3 above can be built using sum-of-products of literals.
Let us define the concrete procedure for obtaining $\be$.
The procedure for $\be'$ is identical, except that the function $f'$ replaces $f$ in the definition.
For every $i = 0, \ldots, 3n-1$, we build the following ``care'' set: \[C_i = \{(a_0, \ldots, a_{k-1}) \in \{0,1\}^k \mid\] \[f(a_0, \ldots, a_{k-1}) = b_0 \ldots b_{3n-1} \; \text{and} \; b_i = 1\}.\]
Then, we can define each $\varphi_i$ as a sum-of-products Boolean algebra expression of the form: 
\[
\varphi_i \equiv \sum_{(a_0, \ldots, a_{k-1}) \in C_i} (g(a_0) \cdots g(a_{k-1}))
\] 
where for $j = 0, \ldots, k-1$,
\[ 
g(a_j) = 
\begin{cases} 
     x_j & a_j = 1, \\
     \bar{x}_j & a_j = 0.
   \end{cases}
\]
If the care set $C_i$ for some $i$ is empty, then $\varphi_i \equiv 0$. 

Let us illustrate the process of obtaining the appropriate Boolean expressions through a simple example.

\begin{example}
    Let $T = \{(2,3,5), (3,3,0), (1,0,9), (8,6,3)\}$, where the optimal values for $\kpid[0][and]$ are $2, 3$ and $5$, respectively. That is, the values we want to protect are those corresponding to the tuple $(2,3,5) \in T$. Then $V = \{0,1,2,3,5,6,8,9\}$ by definition. We interpret the function $\mathit{toBin}$ as the standard conversion from decimal numbers in $\{0,\ldots, 15\}$ to binary numbers of length $n = 4$. For instance, $\mathit{toBin}(2,3,5) = \mathit{toBin}(2) \cdot \mathit{toBin}(3) \cdot \mathit{toBin}(5) = 0010~0011~0101$. We fix $k = 3$.
    We can now build the Boolean expressions needed to apply our strategy.  
   
    First, we take the following arbitrary partition of $\{0,1\}^3$ into $m+1 = |T|$ subsets (see Step~1 above). 
        \begin{align*}
            A_0 = \{000,011\} & \qquad A_1 = \{001,010\}\\
            A_2 = \{100,101\} & \qquad A_3 = \{110,111\}  
        \end{align*}
    
    By definition in Step~2 of our strategy, we get that 
        \begin{align*} 
        f(000) = f(011) = \mathit{toBin}(2,3,5) &= 0010~0011~0101\\
        f(001) = f(010) = \mathit{toBin}(3,3,0) &= 0011~0011~0000\\
        f(100) = f(101) = \mathit{toBin}(1,0,9) &= 0001~0000~1001\\
        f(110) = f(111) = \mathit{toBin}(8,6,3) &= 1000~0110~0011  
        \end{align*}
        
    Choosing $c = 2$, gives us that $f'(x) = f(x)$ whenever $x \notin A_0 \cup A_1$ and $f'(000) = f'(011) = f'(001) = f'(010) = \mathit{toBin}(1,0,9) = 0001~0000~1001$.   
   
   Then, the resulting Boolean expressions for $f$ are:    
   \begin{align*}
       \varphi_0 &\equiv \psi_3 & \varphi_1 &\equiv 0 & \varphi_2 &\equiv \psi_0 + \psi_1\\
       \varphi_3 &\equiv \psi_1 + \psi_2 & \varphi_4 &\equiv 0 & \varphi_5 &\equiv \psi_3\\
       \varphi_6 &\equiv \psi_0 + \psi_1 + \psi_3 & \varphi_7 &\equiv \psi_0 + \psi_1 & \varphi_8 &\equiv \psi_2 \\
       \varphi_9 &\equiv \psi_0 & \varphi_{10} &\equiv \psi_3 & \varphi_{11} &\equiv \psi_0 + \psi_2 + \psi_3
   \end{align*}
   where
   \begin{align*}
       \psi_0 &\equiv \bar{x}_0 \cdot \bar{x}_1 \cdot \bar{x}_2 + \bar{x}_0 \cdot x_1 \cdot x_2 &
       \psi_1 &\equiv \bar{x}_0 \cdot \bar{x}_1 \cdot x_2 + \bar{x}_0 \cdot x_1 \cdot \bar{x}_2\\
       \psi_2 &\equiv x_0 \cdot \bar{x}_1 \cdot \bar{x}_2 + x_0 \cdot \bar{x}_1 \cdot x_2 &
       \psi_3 &\equiv x_0 \cdot x_1 \cdot \bar{x}_2 + x_0 \cdot x_1 \cdot x_2
   \end{align*}
   
   Regarding $f'$, the Boolean expressions $\varphi_0', \varphi_1', \varphi_4', \varphi_5'$ and $\varphi_{10}'$ are exactly as their corresponding Boolean expressions $\varphi_i$ defined above for $f$. The remaining Boolean expressions corresponding to $f'$ are as follows: 
   \begin{align*}
    \varphi'_2 \equiv \varphi'_7 \equiv \varphi'_9 &\equiv 0
    &\varphi'_3 \equiv \varphi'_8 &\equiv \psi_0 + \psi_1 + \psi_2\\
    \varphi'_6 &\equiv \psi_3
    &\varphi'_{11} &\equiv \psi_0 + \psi_1 + \psi_2 + \psi_3
   \end{align*}
\end{example}

Note that the Boolean expressions obtained through the described procedure can be simplified in most cases.
For instance, $\psi_3$ is equivalent to $x_0 \cdot x_1$ as its result is independent of the value assigned to $x_2$.
Moreover, one can use a heuristic logic minimiser such as ESPRESSO\footnote{See \url{https://github.com/scottinet/espresso-logic-minimizer} (based on source code from the University of California, Berkeley).}, which was originally developed at IBM by \citeauthor{espresso} in \citeyear{espresso}~\cite{espresso}, to eliminate redundancies in an automated way.
The result is not guaranteed to be the global minimum but is typically quite close.
The advantage of applying logical minimisation is that the resulting Boolean expressions tend to be much shorter and thus require less evaluation time.

\section{Security Evaluation}\label{sec:secEval}

The attacker's goal is to eliminate the protection that makes the software hardware dependent.
That is, they want to remove the protection so that the software behaves correctly on cloned hardware, not just on the genuine target machine.

Next, we evaluate the security of our approach from the perspective of attackers with increasingly more powerful analysis techniques.
For illustrative purposes, we consider the attacker's options to break the protection applied to a PID controller, which corresponds to discovering the optimal values $\kpid[0][and]$ used by the target MCU.

\subsection{Static Analysis}

The attacker gets access to the protected binary.
We assume they can successfully decompile this binary.
In doing so, they can learn that the values of $\kpid[][and]$ are calculated using the subroutine specified in \autoref{lst:ProtectionKpKiKdValues}.
Moreover, they can also learn the content of the constants $\mathit{hashValue}$, $k$, the Boolean expressions $\be'$, and the static functions $\mathit{toBin}$, $\mathit{toBin}^{-1}$ and $\mathit{eval}$.
That is, we assume the attacker can find out all details except the responses of $\re$ and $\qp$.
This is naturally the case in a static analysis, since the underlying PUF responses can only be obtained by reading the SRAM on the genuine target machine.

With the given information, an attacker can obtain the alternative (suboptimal) values $\kpid[1], \ldots, \kpid[m]$ by calculating $\mathit{toBin}^{-1}(\mathit{eval}(\be', s_i))$ for all $2^k$ binary strings $s_i$ of length $k$ and collecting the results into a set $S$.
However, this does not reveal the optimal values $\kpid[0]$ used on the target MCU by the PID controller, nor does it provide information to logically infer them.
In the unlikely static scenario where the attacker can get hold of the correct Boolean expressions $\be$ by some alternative means, it is still not possible via static analysis to distinguish optimal values $\kpid[0]$ from suboptimal ones, as it would require a dynamic analysis as described in step~(2) in Section~\ref{daClone}.  
Therefore, we can conclude that an adversary performing static analysis alone cannot break the protection.

\subsection{Dynamic Analysis on Cloned MCU}\label{daClone}

The attacker now also has access to a clone of the MCU bound to the protected binary.
Unless the attacker can somehow bypass the protection afforded by $\re$, the program executed on this cloned hardware uses the Boolean expressions $\be'$ to decode the values for $\kpid[][and]$.
Indeed, in our proof of concept the decryption key is, by construction of the PUF (see \autoref{sec:puf}), incorrect.
Independently of the chosen assignment, these alternative Boolean expressions $\be'$ never return the protected values $\kpid[0][and]$. 

Therefore, an attacker has no incentive to perform dynamic analysis on the cloned MCU.
This only changes if the attacker can somehow gain access to the Boolean expressions $\be$.
We stress that this would be an unlikely event considering our proof of concept, as it would require getting a copy of the decryption key, e.g., via industrial espionage.
Assuming the attacker has access to $\be$, then they could perform the following dynamic analysis on the cloned hardware:
\begin{enumerate}[leftmargin=*]
    \item For all $2^k$ binary strings $s_i$ of length $k$, collect the results of $\mathit{toBin}^{-1}(\mathit{eval}(\be, s_i))$ in set $S$ and of $\mathit{toBin}^{-1}(\mathit{eval}(\be', s_i))$ in $S'$. 
    \item Evaluate the results of running the PID algorithm on the cloned hardware for all triples $(\kpid[i])$ in $S \setminus S'$, choosing the one with optimal results.
\end{enumerate}

In the worst case, step~(1) involves evaluating the Boolean expressions with $2^n$ assignments, where $n$ is the number of variables $x$.
However, doing so is possible with relatively low effort.
Logic minimisation techniques can reduce the number of assignments that must be considered for the Boolean variables drastically, unless the size of $S \setminus S'$ is of similar order of magnitude as the number of assignments.
Step (2) is time-consuming, since for each value in $S \setminus S'$ the attacker needs to run the protected program, record the outputs, and identify (with the help of domain knowledge) $\kpid[i][and]$ with the highest efficiency.
We can conclude that the protected values can be identified if the attacker knows the Boolean expressions $\be$.
However, this requires considerable reverse engineering effort.
Whether this effort compensates the gain for the attacker clearly depends on the application at hand.
For PID controllers, this is clearly not the case.

\subsection{Dynamic Analysis on Target MCU}\label{daTarget}

In our threat model, we assume that there is sufficient protection against reading the memory, registers, and execution flow in place.
This implies that debugging and communication interfaces such as JTAG, SWD and UART are disabled or blocked.
In addition, manufacturer-specific functions such as readout protection are activated.
Using an MCU also means that all core components such as the processor and SRAM are housed on a single chip, making it difficult to intercept the communication within; this would require sophisticated physical attacks such as chip scraping.
Under these conditions, even a dynamic analysis on the target hardware cannot directly reveal the correct data values.  
The only relative advantage would be that the attacker could now observe the optimal outputs on the target MCU for different inputs. 
However, they would still not be able to read the PUF responses (nor the values of $\kpid[0][and]$ for that matter) from memory at runtime because the debugging interface was disabled.
This makes the attack described in Section~\ref{daClone} ineffective in recovering the protected data, even though we now have access to the target MCU. 

Considering a man-at-the-end scenario where an attacker has full access to the target MCU and the protected software instance bound to it, they can simply read the protected values $\kpid[0][and]$ from memory.
This scenario could be mitigated by adding another security layer such as leveraging trusted execution environments (TEEs) for handling sensitive data, which we leave for future work.

\section{Proof of Concept}
\label{sec:proof of concept}

In this section, we apply our protection mechanism to a PID controller.
Our proof of concept\footnote{Source code available at \url{https://github.com/COMET-DEPS}} is implemented in C and consists of three parts: (1) a simulated PID controller, (2) an implementation of the SRAM PUF described in \autoref{sec:puf}, and (3) an implementation of the protection mechanism explained in \autoref{sec:mechanism}.

We enrol our SRAM PUF to one of the MCUs, hereinafter referred to as \emph{target MCU}.
We reserve the last 32 bytes of the 512 KiB of SRAM available on the MCU for our PUF.
This allows us to extract a viable \id{} $\key$ with 18 bytes.
We implement the function $\qp$ to return a substring of $\key$.
We use up to 2 bytes returned by $\qp$ as assignment for generating the Boolean expressions $\be$, which we then minimise using ESPRESSO.
We implement $\mathit{encodeExprs}$ to encrypt the minimised expressions with AES-128 using the remaining 16 bytes of $\key$, and $\re$ so that it decrypts them.
For added security, the 16 bytes encryption key should ideally come from a different source than $\qp$, e.g., from a different kind of PUF, see discussion in Section~\ref{sec:discussion}.

At startup of our proof of concept on the target MCU, we reconstruct the enrolled \id{}.
We then extract the key to decrypt the Boolean expressions $\be$.
Subsequently, we execute the simulated PID controller with the constants $\kpid[0][and]$ obtained by evaluating $\be$.
In this case, we observe that the PID controller reliably works as expected.

We then take the view of a potential attacker and copy the proof of concept to an MCU with exactly the same specifications, the \emph{cloned MCU}.
Upon execution of the protected program, we observe that the key obtained from the PUF response does not decrypt the Boolean expressions correctly.
We then see that the Boolean expressions $\be'$ are used as fallback option and evaluated with a Boolean assignment obtained from the (incorrect) PUF response.
The PID controller thus uses values for $\kpid[][and]$ that are not optimal. Although it still reaches a stable point, it does it inefficiently, performing clearly worse than on the target MCU.  
Breaking the encryption to obtain $\be$ does not improve the performance on the cloned MCU, as the PID controller still uses an incorrect assignment for the Boolean expressions (due to an incorrect PUF response), and thus non-optimal $\kpid[][and]$.

\begin{figure}
    \centering
    \includegraphics[width=.9\columnwidth]{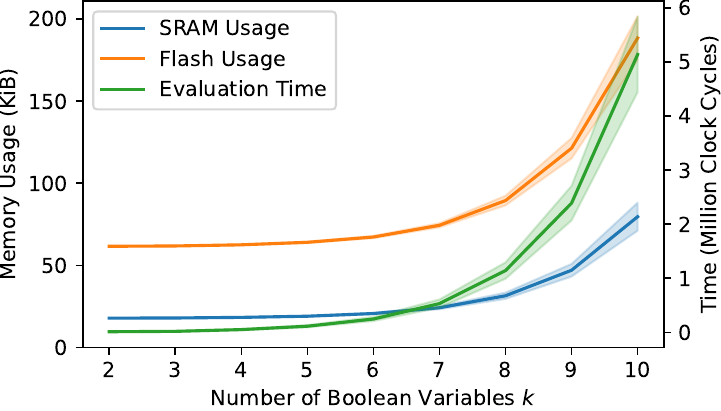}
    \caption{Performance evaluation of our proof of concept for different PUF sizes, the error bands show the standard deviation for different numbers of alternative data values $m$.}
    \label{fig:evaluation}
\end{figure}

To evaluate the performance of our proof of concept, we consider the memory footprint of the protection mechanism and the added time penalty.
For the latter, we measure the evaluation time of the resulting Boolean expressions for different numbers $k$ of Boolean variables as well as different numbers of alternative data values $m$.
The results are shown in \autoref{fig:evaluation}.
We calculate memory usage with \emph{PlatformIO}, which adds the size of the \texttt{.text} and \texttt{.data} sections of the binary program for the flash usage, and the size of the \texttt{.bss} and \texttt{.data} sections for the RAM usage\footnote{For details see \url{https://docs.platformio.org/en/latest/faq/program-memory-usage.html}}.
With larger values for $k$, the size of the Boolean expressions $\be$ increases exponentially (as expected), whereas $m$ has only limited influence.
With an exponential growth of the size of the Boolean expressions, the SRAM and flash usage also increase exponentially, as well as the startup time of the proof of concept due to the longer evaluation times.
For $k=6$, the average evaluation time is 1.14~ms at a CPU frequency of 216~MHz, for $k=10$ it is 23.79~ms  and thus still acceptable.

\section{Discussion}
\label{sec:discussion}

A main advantage of using Boolean expressions to encode the protected values instead of just encrypting them, is that this enables the implementation of dual layer protection.
In our proof of concept, we simply use two different PUF responses, one to decrypt the Boolean expressions and another one to obtain the correct assignment for the Boolean variables in those expressions.
Instead, we could consider a different source for the key needed to decrypt the Boolean expressions or even replace encryption with GlueZilla's approach~\cite{gluezilla}, provided the device has DRAM.
If an attacker can break the encryption, they still cannot access the target values unless they can also access the required Boolean assignment.
Nevertheless, the attacker could pinpoint a set of data values that contains the secret ones. 
However, they would not be able to distinguish the correct (secret) values from the suboptimal ones unless the behaviour of the machine in each case is evaluated by a domain expert. 
Finally, if an attacker cannot decrypt the Boolean expressions but instead can get the right assignment, the fallback Boolean expressions still do \emph{not} return the secret data values.

On legacy hardware, we could use our mechanism to implement a secure update procedure, which transmits encrypted data to a device, and the best performance can only be achieved based on the correct PUF response.
If we consider newer MCUs that have additional security features such as TEEs, our dual layer approach could be applied to receiving sensitive data in encrypted form and transmitting it decrypted into the secure enclave, which then evaluates the Boolean expressions, thus representing an additional protection layer.

Essential future work should consider how to better handle the secret data values
later at runtime, to prevent that even a memory dump does not result in these values being revealed.
A possible approach could be to keep these data values in registers.
Another possibility and a further security layer would be to take advantage of the mentioned TEEs.

A limiting factor discussed in \autoref{sec:proof of concept} is the exponential growth in size of the resulting Boolean expressions in relation to the considered number of Boolean variables.
Nevertheless, our protection mechanism increases memory usage only during startups and the time penalty is within reasonable limits for the target applications.

\section{Related Work}
\label{sec:related-work}

\citeauthor{Zhang2015} proposed a lock mechanism to protect IP in FPGAs~\cite{Zhang2015}.
The security of their method has been the subject of some controversy~\cite{Zhang2015comment,Zhang2015rebuttal}.
They use a PUF response to control an added finite-state machine (FSM).
Only if this FSM reaches an unlocking state, the original code is executed, whereas our approach allows to execute suboptimal code instead.
Their mechanism is specific to FPGA devices and not directly applicable to our use case, i.e., to the protection of data in embedded software in general. 

Our work is partially inspired by GlueZilla~\cite{gluezilla}, a system that binds software to hardware through user-space Rowhammer PUFs.
This system transforms binary programs in memory at runtime so that they only exhibit their correct behaviour on the genuine target machine to which they were bound to at compile time.
When run on any other machine, the programs will exhibit a different behaviour.
This method cannot be applied in our proof of concept since Rowhammer PUFs are not realisable on SRAM, they require DRAM.
Moreover, the way in which this protection mechanism works is intrinsically different to ours, as it is intended to protect program logic rather than program data.

Likewise, an approach proposed by \citeauthor{traffic-lights} protects software logic rather than its data~\cite{traffic-lights}.
This method can be applied to protect embedded software, but only if the logic of the program is sufficiently complex and corresponds to a control state algorithm.
This is not the case, for instance, for PID controllers as those considered in our proof of concept, where the critical IP to be protected is program data and the correct program logic is well known.

None of the mentioned related works uses a dual layer protection as proposed here.

\section{Conclusion}\label{sec:concl}

In this paper, we introduce a novel protection mechanism to safeguard intellectual property in the form of data in embedded software.
Our dual layer approach consists of encoding the data to be protected as Boolean expressions, and the protection of these expressions.
This not only adds two layers of security, but also allows using two different sources to recover the expressions and the assigned values.
In our proof of concept, we identify the genuine target MCU by leveraging physical properties of its built-in SRAM, making our mechanism work with legacy hardware.
Only if both the key used to decrypt the Boolean expressions and their assigned values extracted from an MCU’s SRAM match the key and values expected from the target MCU, the protected data is recovered correctly and the software using it is executed as expected.
On any other MCU, we use default Boolean expressions that are guaranteed to deliver safe values independently of the Boolean assignment when evaluated.
Thus, the embedded software we protect can operate safely on any device, yet suboptimally.
Bypassing our protection mechanism is costly and complex, requiring dynamic analysis of the software on the target MCU to uncover the secret data, making it a robust solution for safeguarding IP in control systems.

\begin{acks}
The research reported in this paper has been funded by the Federal Ministry for Climate Action, Environment, Energy, Mobility, Innovation and Technology (BMK), the Federal Ministry for Labour and Economy (BMAW), and the State of Upper Austria in the frame of the COMET Module Dependable Production Environments with Software Security (DEPS) (FFG grant no. 888338) and the SCCH competence center INTEGRATE (FFG grant no. 892418) within the COMET - Competence Centers for Excellent Technologies Programme managed by Austrian Research Promotion Agency FFG.
\end{acks}

\bibliographystyle{ACM-Reference-Format}
\bibliography{bibl} 

\end{document}